\documentclass[twocolumn,showpacs,preprintnumbers,amsmath,amssymb]{revtex4}

\usepackage{graphicx}
\usepackage{dcolumn}
\usepackage{bm}

\usepackage{axodraw}
\usepackage{color}

\def\be{\begin{equation}}
\def\ee{\end{equation}}
\def\bea{\begin{eqnarray}}
\def\eea{\end{eqnarray}}

\begin{document}

\preprint{APS/123-QED}

\title{Thermal Corrections to $\pi$-$\pi$ Scattering Lengths in the Linear Sigma Model}

\author{M. Loewe}
 \email{mloewe@fis.puc.cl}
{.}
\author{C. Martinez-Villalobos}%
 \email{cxmartin@uc.cl}
\affiliation{Facultad de F\'\i sica, Pontificia Universidad
Cat\'olica de Chile,\\ Casilla 306, Santiago 22, Chile. }


\date{\today}

\begin{abstract}
In this article we address the problem of getting the temperature
dependence of the $\pi$-$\pi$ scattering lengths in the frame of
the linear sigma model. Using the real time formalism, we
calculate all the relevant one loop diagrams. The temperature
corrections are only considered in the pion sector, due to the
Boltzmann suppression for heavier fields like the sigma meson.
From this analysis we obtain the thermal behavior of the s-waves
scattering lengths $a_0^0$ and $a_0^2$ associated to isospin $I=0$
and $I=2$, respectively. If we normalize with the zero temperature
value it turns out that $\frac{a_0^0(T)}{a_0^0}$ grows with
temperature, whereas the opposite occurs with
$\frac{a_0^2(T)}{a_0^2}$. Finally we compare our results with
other determinations of the scattering lengths based on the Nambu-Jona-Lasinio model and chiral perturbation theory.\\

\end{abstract}

\maketitle

\section{Introduction}

The discussion of the thermal evolution of $\pi$-$\pi$ scattering
lengths turns out to be a relevant problem in the context of heavy
ion collisions. In fact, we know that in the central rapidity
region, precisely where the quark-gluon plasma is expected to be
created, a big amount of thermalized pions will be produced. Those
pions will interact among themselves and the $\pi$-$\pi$
scattering lengths are crucial parameters in order to describe the
scattering amplitudes. At zero temperature the $\pi$-$\pi$
scattering lengths were first measured by Rosellet et al
\cite{Rosellet}. A recent review about the present status of the
experimental measurements of $\pi$-$\pi$ scattering lengths and
their comparison with different
theoretical approaches can be found in \cite{Urets}.\\
In this article we will present a detailed calculation of the
thermal corrections to the $\pi$-$\pi$ scattering lengths in the
frame of the linear sigma model \cite{Gell-Mann}. As it is well
known, the linear sigma model is an effective, renormalizable
\cite{Lee}, low energy description of hadron dynamics. Our
calculations will be done using the real time
formalism at the one loop level.\\


\section{Linear sigma model and $\pi$-$\pi$ scattering}

The linear sigma model in the phase where the chiral symmetry is
broken is given by the lagrangian below, where
$v=\langle\sigma\rangle$ is the vacuum expectation value of the
scalar field $\sigma$. The idea is to define a new field $s$ such
that $\sigma = s+v$. Obviously $\langle s\rangle=0$. $\psi$
corresponds to an isospin doublet associated to the nucleons,
$\vec{\pi}$ denotes the pion isotriplet field and $c\sigma$ is the
term that breaks explicitly the $SU(2)\times SU(2)$ chiral
symmetry. $\epsilon$ is a small dimensionless parameter. It is
interesting to remark that all fields in the model have masses
determined by $v$. In fact, the following relations are valid:
$m_{\psi}=gv$, $m_{\pi}^2=\mu^2+\lambda^2v^2$ and
$m_{\sigma}^2=\mu^2+3\lambda^2v^2$. Perturbation theory at the
tree level allows us to identify the pion decay constants as
$f_{\pi}=v$. This model has been considered in the context of
finite temperature by several authors, discussing the thermal
evolution of masses, $f_\pi(T)$, the effective potential, etc
\cite{Loewe}, \cite{Larsen}, \cite{Bilic}, \cite{Petropolus}.\\

\begin{align}
\mathcal{L}&=\bar{\psi}[i\gamma^{\mu}\partial_{\mu}-m_{\psi}-g(s+i\vec{\pi}\cdot\vec{\tau}\gamma_{5})]\psi+\frac{1}{2}[(\partial\vec{\pi})^2+m_{\pi}^2\vec{\pi}^2]\nonumber\\
&+\frac{1}{2}[(\partial\sigma)^2+m_{\sigma}^2 s^2]-\lambda^2
vs(s^2+\vec{\pi}^2)-\frac{\lambda^2}{4}(s^2+\vec{\pi}^2)^2\nonumber\\
&+(\varepsilon c-vm_{\pi}^2)s
\end{align}

Since our idea is to use the linear sigma model for calculating
$\pi$-$\pi$ scattering lengths, let us remind briefly the
formalism. The scattering amplitude has the general form

\begin{align}
T_{\alpha\beta;\delta\gamma}&=A(s,t,u)\delta_{\alpha\beta}\delta_{\delta\gamma}+A(t,s,u)\delta_{\alpha\gamma}\delta_{\beta\delta}\nonumber\\
&+A(u,t,s)\delta_{\alpha\delta}\delta_{\beta\gamma}.
\end{align}

\noindent where the $\alpha$, $\beta$, $\gamma$, $\delta$ denote
isospin components.

By using appropriate projection operators, it is possible to find
the following isospin dependent scattering amplitudes



\begin{align}
T^{0}&=3A(s,t,u)+A(t,s,u)+A(u,t,s)\\
T^{1}&=A(t,s,u)-A(u,t,s)\\
T^{2}&=A(t,s,u)+A(u,t,s),
\end{align}

\noindent where $T^I$ denotes a scattering amplitude in a given isospin channel.\\

As it is well known \cite{Collins}, the isospin dependent
scattering amplitude can be expanded in partial waves $T_l^I$.

\begin{equation}
T_{\ell}^{I}(s)=\frac{1}{64\pi}\int_{-1}^{1}d(cos\theta)P_{\ell}(cos\theta)T^{I}(s,t,u)
\end{equation}

Below the inelastic threshold the partial scattering amplitudes
can be parametrized as \cite{Gasser}.


\begin{equation}
T_{\ell}^{I}=(\frac{s}{s-4m\pi^2})^{\frac{1}{2}}\frac{1}{2i}(e^{2i\delta_{\ell}^{I}(s)}-1),
\end{equation}

\noindent where $\delta_{\ell}$ is a phase-shift in the $\ell$
channel. The scattering lengths are important parameters in order
to describe low energy interactions. In fact, our last expression
can be expanded according to

\begin{equation}
\Re(T_{\ell}^{I})=(\frac{p^{2}}{m_{\pi}^{2}})^{\ell}(a_{\ell}^{I}+\frac{p^2}{m_{\pi}^{2}}b_{\ell}^{I}+...)
\end{equation}

The parameters $a_{\ell}^{I}$ and $b_{\ell}^{I}$ are the
scattering lengths and scattering slopes, respectively. In
general, the scattering lengths obey $|a_{0}|>|a_{1}|>|a_{2}|...$.
If we are only interested in the scattering lengths $a_0^I$, it is
enough to calculate the scattering amplitude $T^I$ in the static
limit, i.e. when $s \to 4m_\pi^2$, $t\to 0$ and $u\to 0$

\begin{equation}
a_{0}^{I}=\frac{1}{32\pi}T^{I}(s \to 4m_{\pi}^2,t\to 0, u\to0)
\end{equation}

\section{Pion-Pion Scattering Amplitudes}

The diagrams shown in figure 1, where the continuum line denotes a
pion, and the dashed line a sigma meson, contribute to the
$\pi$-$\pi$ scattering amplitude. The diagram with a sigma
exchanged meson has
to be considered also in the crossed $t$ and $u$ channels.\\
From these diagrams it is possible to get the results shown in
table I. The isospin dependent scattering amplitudes at the tree
level have the form

\begin{align}
T^{0}(s,t,u)&=-10\lambda^2-\frac{12\lambda^{4}v^{2}}{s-m_{\sigma}^2}-\frac{4\lambda^{4}v^{2}}{t-m_{\sigma}^2}-\frac{4\lambda^{4}v^{2}}{u-m_{\sigma}^2}\\
T^{1}(s,t,u)&=\frac{4\lambda^{4}v^{2}}{u-m_{\sigma}^2}-\frac{4\lambda^{4}v^{2}}{t-m_{\sigma}^2}\\
T^{2}(s,t,u)&=-4\lambda^2-\frac{4\lambda^{4}v^{2}}{t-m_{\sigma}^2}-\frac{4\lambda^{4}v^{2}}{u-m_{\sigma}^2}.
\end{align}

\begin{figure}[h!]
\begin{center}
\includegraphics[scale=.9]{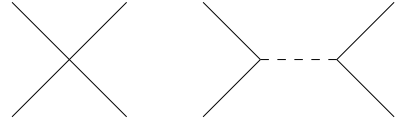}
\caption{Tree level diagrams}
\end{center}
\end{figure}






\begin{table}
\begin{tabular}{|l|p{2.5cm}|p{2.5cm}|p{2.5cm}|}
\hline
 & Experimental Results & Chiral Perturbation Theory & Linear
 Sigma Model
 \\ \hline
$a_0^0$& $0.26\pm 0.05$&$\frac{7m_\pi^2}{32\pi
f_\pi^2}=0.16$&$\frac{10m_\pi^2}{32\pi f_\pi^2}=0.22$\\ \hline
$b_0^0$& $0.25\pm 0.03$&$\frac{m_\pi^2}{4\pi
f_\pi^2}=0.18$&$\frac{49m_\pi^2}{128\pi f_\pi^2}=0.27$\\ \hline
$a_0^2$& $-0.028\pm 0.012$&$\frac{-m_\pi^2}{16\pi
f_\pi^2}=-0.044$&$\frac{-m_\pi^2}{16\pi f_\pi^2}=-0.044$\\ \hline
$b_0^2$& $-0.082\pm 0.008$&$\frac{-m_\pi^2}{8\pi
f_\pi^2}=-0.089$&$\frac{-m_\pi^2}{8\pi f_\pi^2}=-0.089$\\ \hline
$a_1^1$& $0.038\pm 0.002$&$\frac{m_\pi^2}{24\pi
f_\pi^2}=0.030$&$\frac{m_\pi^2}{24\pi f_\pi^2}=0.030$\\ \hline
$b_1^1$& $-$&$0$&$\frac{m_\pi^2}{48\pi f_\pi^2}=0.015$\\

\hline

  \hline
\end{tabular}
\caption{Comparison between the experimental values
\cite{Rosellet}, first order prediction from chiral perturbation
theory \cite{Weinberg} and our results at the tree level.}
\end{table}

 Note that, the linear sigma model is in a better agreement with the
experimental results than first order
chiral perturbation theory.\\


\section{One loop thermal corrections for scattering lengths}

For our calculation of the thermal corrections to the scattering
lengths, we will use the real time formalism. At the one loop
level it is enough to use the Dolan-Jackiw propagators
\cite{Dolan:1973qd}. A general review about the real time
formalism, beyond the one loop level can be found in
\cite{Das:1997}, \cite{LeBellac:1996}. In our case the most
relevant thermal contributions will be related to the pion sector,
due to the Boltzmann suppression in the case of the sigma meson
and/or nucleons. Therefore, for the pion propagators we will use

\begin{align}
\Delta(k_{0},\vec{k},m_{\pi})\delta_{\alpha\beta}&=
(\frac{i}{k_0^2-\vec{k}^2-m_{\pi}^2+i\epsilon}+2\pi
n_{B}(|k_{0}|)\nonumber\\
&\delta(k_0^2-\vec{k}^2-m_{\pi}^2))\delta_{\alpha\beta},
\end{align}

\noindent where $n_B(x)=\frac{1}{e^{\beta x}-1}$ is the
Bose-Einstein distribution. For the $\sigma$ meson propagator, we
will restrict ourselves to the $T=0$ case

\begin{align}
\Delta(k_{0},\vec{k},m_{\sigma})&=\frac{i}{k_0^2-\vec{k}^2-m_{\sigma}^2+i\epsilon}.
\end{align}

There are many diagrams that contribute to the pion-pion
scattering amplitude at the one loop level. For each one of these
diagrams we have to add also the corresponding crossed t and u
channel diagrams. In figure 2 we have shown only the s channel
contribution.

We will give the analytic expression only for the first two
diagrams (a,b) shown in figure 2. It should be noticed that, when
it corresponds, symmetry factors, isospin index contractions and
multiplicity factors should be included.

\begin{figure}[h!]
\begin{center}
\includegraphics[scale=.9]{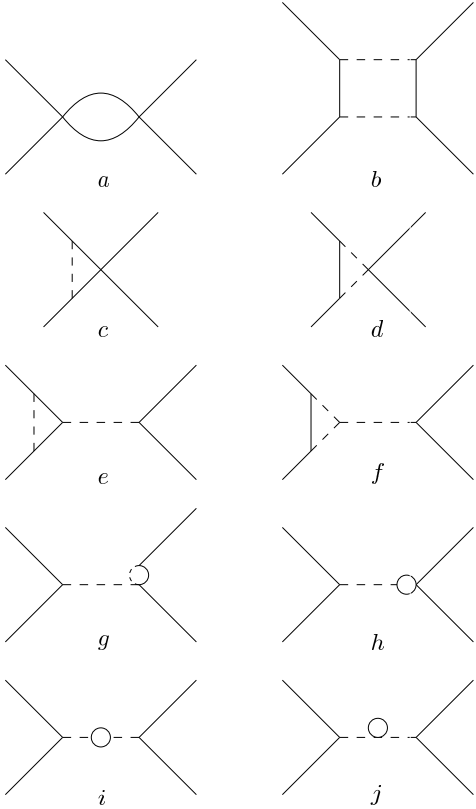}
\caption{Relevant one loop diagrams}
\end{center}
\end{figure}

\begin{align}
i\emph{$M_a$}&=-2\lambda^4(7\delta_{\alpha\beta}\delta_{\gamma\delta}+2\delta_{\alpha\gamma}\delta_{\beta\delta}
+2\delta_{\alpha\delta}\delta_{\beta\gamma})\nonumber\\
&\times\int\frac{d^4
k}{(2\pi)^4}\Delta(k_{0},\vec{k},m_{\pi})\Delta(k_{0}-2m_{\pi},\vec{k},m_{\pi})
\end{align}

\begin{align}
i\emph{$M_b$}=&16\lambda^8
v^4(\frac{i}{4m_\pi^2-m_\sigma^2})\delta_{\alpha\beta}\delta_{\gamma\delta}\int\frac{d^4
k}{(2\pi)^4}[\Delta(k_{0},\vec{k},m_{\sigma})&\nonumber\\
&\Delta(k_{0}+m_{\pi},\vec{k},m_{\pi})\Delta(k_{0}-m_{\pi},\vec{k},m_{\pi})]
\end{align}



In some of these diagrams we have to deal with terms $\delta^2 (k)$,
and $\frac{\delta(k)}{k}$. In those cases the following identity is
useful \cite{Umezawa}

\begin{equation}
\frac{1}{N!}(\frac{i\partial}{\partial
m^{2}})^{n}\Delta=\Delta^{n+1}.
\end{equation}

Actually this expression can be proved in the full matrix
formalism of Thermofield Dynamics. However, for some diagrams, we
checked our results
using also the Matsubara (imaginary time) formalism. \\

When dealing with integrals of the following shape,

\begin{align}
&I=2\pi i\int dk_0 F(k_0)(\frac{\partial}{\partial
m_{\pi}^{2}}[\frac{n_{B}(|k_0|)}{2w_k}](\delta(k_0-w_k)
\\ \nonumber
&+\delta(k_0+w_k))
+(\frac{n_{B}(|k_0|)}{2w_k})\frac{\partial}{\partial
m_{\pi}^{2}}[\delta(k_0-w_k)+\delta(k_0+w_k)])
\end{align}

where $F(k_0)$ is an arbitrary function and
$w_k=\sqrt{\vec{k}^{2}+m_{\pi}^{2}}$, we can integrate by parts
getting $I=I_{w_{k}}+I_{-w_{k}}$, where


\begin{align}
I_{w_{k }}&=2\pi\lim_{k_0\to
w_k}(\frac{i(\frac{dF(k_0)}{dk_0})n_{B}(|k_0|)}{4w_k^2}
\nonumber\\
&-i\frac{F(k_0)sgn(k_0)(csch(\frac{|k_0|}{2T}))^{2}}{16w_k^2
T}-i\frac{F(w_k)n_B(w_k)}{4w_k^3})
\end{align}

and

\begin{align}
I_{-w_{k}}&=2\pi\lim_{k_0\to
-w_k}(\frac{-i(\frac{dF(k_0)}{dk_0})n_{B}(|k_0|)}{4w_k^2}
\nonumber\\
&+i\frac{F(k_0)sgn(k_0)(csch(\frac{|k_0|}{2T}))^{2}}{16w_k^2
T}-i\frac{F(-w_k)n_B(w_k)}{4w_k^3}).
\end{align}

 After taking all the diagrams into account, we find the following expression for the one loop
thermal corrections to the $\pi$-$\pi$ scattering amplitudes

\begin{align}
&A(s,t,u)_T=-\lambda^4(14+80\lambda^2
v^2(\frac{1}{4m_\pi^2-m_\sigma^2})\nonumber\\
&+48\lambda^4 v^4(\frac{1}{4m_\pi^2-m_\sigma^2})^2)A(w_k,T)
-8\lambda^4 B(w_k,T)\nonumber\\
&-12\lambda^6
v^2(\frac{1}{4m_\pi^2-m_\sigma^2})^2
C(w_k,T)\nonumber\\
&+\lambda^6 (16v^2+32\lambda^2
v^4(\frac{1}{4m_\pi^2-m_\sigma^2}))\nonumber\\
&\times[D^{1}(m_{\pi},m_{\sigma},w_k,T)+D^{1}(-m_{\pi},m_{\sigma},w_k,T)]\nonumber\\
&+32\lambda^6 v^2[E_{0}^{1}(m_\pi,m_\sigma,w_k,T)+E_{0}^{1}(-m_\pi,m_\sigma,w_k,T)]\nonumber\\
&+32\lambda^8 v^4
[E_{1}^{1}(m_\pi,m_\sigma,w_k,T)+E_{1}^{1}(-m_\pi,m_\sigma,w_k,T)]\nonumber\\
&+8\lambda^6 v^2(\frac{1}{4m_\pi^2-m_\sigma^2})
\times[F_{0}^{1}(m_\pi,m_\sigma,w_k,T)\nonumber\\
&+F_{0}^{1}(-m_\pi,m_\sigma,w_k,T)]+\lambda^6 [96\lambda^2
v^4(\frac{1}{4m_\pi^2-m_\sigma^2})\nonumber\\
&+16v^2] F_{1}^{1}(m_\pi,m_\sigma,w_k,T),
\end{align}

\begin{align}
&A(t,s,u)_T=-4\lambda^4 A(w_k,T)-\lambda^4(11+80\lambda^2
v^2(\frac{1}{-m_\sigma^2})\nonumber\\
&+48\lambda^4 v^4(\frac{1}{-m_\sigma^2})^2)B(w_k,T)+4\lambda^6
v^2(\frac{1}{-m_\sigma^2})\nonumber\\
&\times(\frac{4}{m_\sigma^2-m_\pi^2}+\frac{3}{m_\sigma^2})
C(w_k,T)\nonumber\\
&+16\lambda^6
v^2[D^{1}(m_\pi,m_\sigma,w_k,T)+D^{1}(-m_\pi,m_\sigma,w_k,T)]\nonumber\\
&+16\lambda^8 v^4
[D^{2}(m_\pi,m_\sigma,w_k,T)+D^{2}(-m_\pi,m_\sigma,w_k,T)]\nonumber\\
&+\lambda^6 (32v^2+32\lambda^2
v^4(\frac{1}{-m_\sigma^2}))\nonumber\\
&\times[E^{1}_{0}(m_\pi,m_\sigma,w_k,T)+E^{1}_{0}(-m_\pi,m_\sigma,w_k,T)] \nonumber\\
&+16\lambda^8 v^4 [E^{2}_{0}(m_\pi,m_\sigma,w_k,T)+E^{2}_{0}(-m_\pi,m_\sigma,w_k,T)] \nonumber\\
&+\lambda^6 (48\lambda^2
v^4(\frac{1}{-m_\sigma^2})+8v^2)[F^{2}_{0}(m_\pi,m_\sigma,w_k,T)\nonumber\\
&+F^{2}_{0}(-m_\pi,m_\sigma,w_k,T)]\nonumber\\
&=A(u,t,s)_T,
\end{align}

where we have introduced the following definitions


\begin{equation}
A(w_k,T)=\frac{-1}{(2\pi)^2}\int d|\vec{k}|\frac{n_B(w_k)}{w_k},
\end{equation}

\begin{equation}
B(w_k,T)=-\int \frac{d^3
k}{(2\pi)^3}(\frac{n_B(w_k)}{2w_k^3}+\frac{(csch(\frac{w_k}{2T}))^2}{8Tw_k^2}),
\end{equation}

\begin{equation}
C(w_k,T)=\int \frac{d^3k}{(2\pi)^3}\frac{n_B(w_k)}{w_k},
\end{equation}

\begin{align}
D^{n}(x,y,w_k,T)&=-\int \frac{d^3
k}{(2\pi)^3}\frac{n_B(w_k)}{w_k}\nonumber\\
&\times [\frac{1}{f^{n}(x,y)}(\frac{1}{4x^2+4xw_k})]
\end{align}

\begin{align}
&E^{n}_{m}(x,y,w_k,T)=\int \frac{d^3
k}{(2\pi)^3}[\frac{(x+w_k)n_B(w_k)}{2w_k^2f^{n+1}(x,y)f^{m}(-x,y)}\nonumber\\
&(\frac{(csch(\frac{w_k}{2T}))^2}{16Tw_k^2}+\frac{n_B(w_k)}{4w_{k}^3})(\frac{1}{f^{n}(x,y)f^{m}(-x,y)})]
\end{align}

\begin{equation}
F^{n}_{m}(x,y,w_k,T)=-\int\frac{d^3k}{(2\pi)^3}\frac{n_B(w_k)}{w_k}[\frac{1}{f^{n}(x,y)f^{m}(-x,y)}],
\end{equation}

and where

\begin{equation}
f^{n}(x,y)=(2x^2+2xw_{k}-y^2)^n.
\end{equation}

\section{Numerical results}

The thermal corrections must be calculated numerically. The
different parameters in our expressions are renormalized at $T=0$
, since thermal corrections do not add new ultraviolet
divergencies. The linear sigma model, excluding the nucleons, has
three parameters: $m_{\pi}^{2}$, $f_{\pi}$ and $\lambda^{2}$. The
first two parameters, $m_{\pi}^{2}$ and $f_{\pi}$, are given by
experiments and the third one is a free parameter. Notice that
$f_\pi$ is related to the vacuum expectation value $v$. In fact,
at the tree level $f_\pi=v$. The three parameters are not
independent. If instead of $f_\pi$ we use the vacuum expectation
value $v$ and consider a mass of the sigma meson $m_{\sigma}=700$
MeV, we have $\lambda^2=7$, $v=90$ MeV; if $\lambda^2=5.6$,
$v=120$ MeV \cite{Basdevant}. We know, however, that the mass of
the sigma meson is about $m_\sigma=550$ MeV \cite{Aitala}.
Therefore, we need to find new values for $\lambda$ and $v$
associated to the new lower mass of the sigma meson. We found
$\lambda^2=4.26$ and $v=89$ MeV, following the philosophy
presented in \cite{Basdevant}. The authors used the Pad\'e
approximant method, but they also suggested an analytic approach
in order to find the variation of one parameter if the other two
change as a consequence of one loop corrections. The values given
above for
$\lambda^2$ and $v$ were found following this procedure.\\
The scattering lengths, including our thermal corrections are
given by

\begin{align}
a_0^0(T)&=0.217+\frac{3A(s,t,u)_T+2A(t,s,u)_T}{32\pi}\\
a_0^2(T)&=-0.041+\frac{2A(t,s,u)_T}{32\pi}.
\end{align}

\begin{figure}[h!]
\begin{center}
\includegraphics[scale=.9]{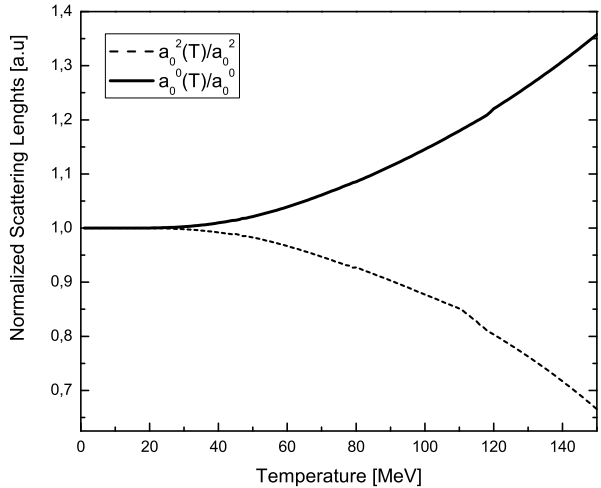}
\caption{Scattering lengths normalized to $T=0$}
\end{center}
\end{figure}

The behavior of the normalized scattering lengths
$\frac{a_0^0(T)}{a_0^0}$ and $\frac{a_0^2(T)}{a_0^2}$ are shown in
figure 3. Notice that $a_0^1(T)$ vanishes identically. We
normalized with the zero temperature values of the scattering
lengths, according to a two loop perturbation calculation
\cite{Bijnens}. It is interesting to remark that $a_0^0(T)$ grows
with temperature and the same occurs with the $a_0^2(T)$. Similar
calculations have been done in the context of the
Nambu-Jona-Lasinio model \cite{Quack} and in the frame of chiral
perturbation theory \cite{Kaiser}. The results from the
Nambu-Jona-Lasinio analysis do not agree with our conclusions. In
this approach both scattering lengths remain almost constant, but
with the same growing tendency. The results from chiral
perturbation theory agree quite well with our conclusions for
$a_0^2(T)$. Nevertheless, $a_0^0(T)$ is almost constant in this
approach. In \cite{Schaefer}, a different determination of
$a_{0}^{0}(T)$ based on the Heat Kernel expansion technique
applied to the linear sigma model, is presented. We agree with
their results.\\

\section*{ACKNOWLEDGMENTS}

We acknowledge support from Fondecyt under grant Nr. 1051067. M.L.
acknowledges also support from the Centro de Estudios Subat\'omicos.
We also would like to thank Prof. A. Ayala and Prof. A. Das for
helpful correspondence. We thank several discussions with Jorge
Ruiz.


\end{document}